\documentstyle[natbib,amssymb,graphicx]{aipproc}

\graphicspath{{/home/katja/texte/conferences/compton5_sept_99/poster/}}

\newlength{\figurewidth}
\begin{document}

\title{Monitoring the Short-Term Variability of Cyg X-1: Spectra and Timing}

\author{K.~Pottschmidt$^*$, J.~Wilms$^*$,
  R.~Staubert$^*$, M.A.~Nowak$^\dagger$, J.B.~Dove$^{\ddagger,\|}$,
  W.A.~Heindl$^{\P}$, D.M.~Smith$^{\S}$}

\address{$^{*}$ Institut f\"ur Astronomie und Astrophysik,
  Waldh\"auser Str. 64, D-72076 T\"ubingen, Germany, \\
  $^{\dagger}$ JILA,
  University of Colorado, Boulder, CO 80309-440, U.S.A., \\
  $^{\ddagger}$
  CASA, University of Colorado, Boulder, CO 80309-389, U.S.A.,\\
  $^{\|}$ Dept.\ of Physics, Metropolitan State College of Denver, Denver,
  CO 80217-3362, U.S.A.,\\
  $^{\P}$ CASS,
  University of California San Diego, La Jolla, CA 92093, U.S.A.,\\
  $^{\S}$ SSL, University of California Berkeley, Berkeley, CA 94720,
  U.S.A.}

\maketitle

\begin{abstract}
  We present first results from the spectral and temporal analysis of an RXTE
  monitoring campaign of the black hole candidate Cygnus X-1 in 1999.
  The timing properties of this hard state black hole show considerable
  variability, even though the state does not change. This has previously
  been noted for the power spectral density, but is probably even more
  pronounced for the time lags. From an analysis of four monitoring
  observations of \mbox{Cyg X-1}, separated by 2 weeks from each other, we find
  that a shortening of the time lags is associated with a hardening of the
  X-ray spectrum, as well as with a longer characteristic ``shot time scale''.
  We briefly discuss possible physical/geometrical reasons for this
  variability of the hard state properties. 
\end{abstract}

\section*{Introduction}

For stellar black hole candidates, several distinct states can be
identified that differ in their general spectral and temporal properties.
Based mainly on spectral arguments these states have been associated with
different accretion rates and different geometries of the accretion flow
\citep[e.g.,][]{esin:98a,nowak:95a}. With broad band instruments like the
Rossi X-ray Timing Explorer (RXTE) it is possible to study the states with
high time resolution over a time base of several years. The focus of this
work lies on parameters and functions characterizing the short-term
variability ($<1000$\,s) of the canonical black hole candidate \mbox{Cyg
  X-1} and their stability in the hard state.

\begin{figure}
\centerline{\includegraphics[width=0.6\textwidth]{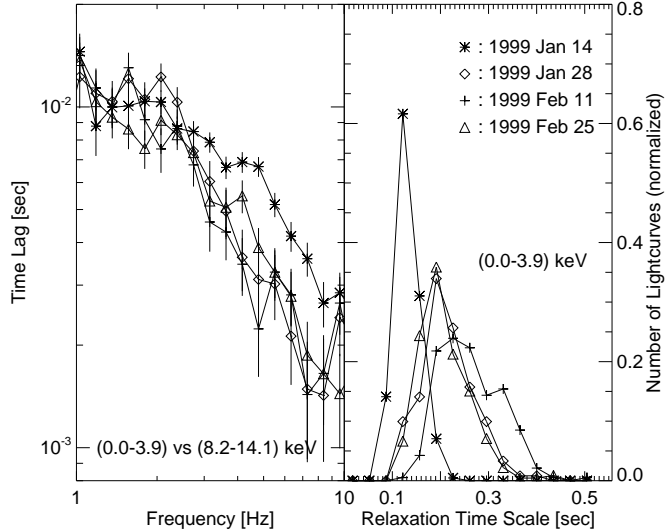}}
\caption{Comparison of four consecutive RXTE observations of 1999 January and
  February, spaced by $\sim$14\,d each. \textbf{Left:} Time lags as a
  function of Fourier frequency. The time lag in the 1 to 10\,Hz band
  changes by a factor of three over the course of a month. Note the
  logarithmic $y$-axis! \textbf{Right:} Distribution of the relaxation time
  scale $\tau$ found from short (32\,s long) time segments for these
  observations. Observations with shorter time lags appear to have larger
  $\tau$.}
\label{fig:zrmcompare}
\end{figure}

In 1996 and 1997 observations of \mbox{Cyg X-1} with the pointed RXTE
instruments were not performed regularly and mainly concentrated on the
$\sim$3 month long soft state in 1996 \citep[see, e.g.,][]{cui:98c}. We
initiated a monitoring campaign of the hard state in 1998 (3\,ksec
exposures), which we expanded to 10\,ksec exposures in 1999 in order to
allow the calculation of Fourier frequency dependent time lags with
sufficient accuracy (Fig.~\ref{fig:zrmcompare}). Additionally, the RXTE
observations are accompanied by simultaneous radio pointings. The aim of
this campaign is to address fundamental questions such as the cause of the
long term flux variability in the hard state, namely the 150\,d periodic
behavior seen in the RXTE All Sky Monitor and in the radio flux
\citep[][Hjellming, priv.\ comm.]{pooley:98a}. A precessing, interacting
disk-jet system has been suggested as one possible explanation for this
hard state cycle \citep{brocksopp:99b}.

We have performed spectral and/or temporal analyses on $\sim$ 30\% of the
available public data measured before 1999. In addition, we have analyzed
those of our 2\,weekly observations in 1999 that were scheduled before the
gain change of the RXTE Proportional Counter Array (PCA) in 1999 March (for
data after the gain change the calibration and background models are still
uncertain). In this paper we present first results from these monitoring
observations.  Using the \texttt{ftools} 4.2, we extracted PCA spectra and
high (2\,ms) time resolution PCA lightcurves. We computed periodograms for
several energy bands, as well as the time lags, and the coherence function
between these energy bands \citep{nowak:98a}. In addition, we use the
linear state space model (LSSM) to model the light curves in the time
domain. This method allows one to derive a characteristic time scale,
$\tau$, that can explain the dynamics of the lightcurve. $\tau$ can be
interpreted in terms of a shot noise relaxation time scale, but note that
LSSMs only need a single time scale to provide a good fit of the lightcurve
\citep[see ][for an application of the LSSM to EXOSAT data from
Cyg~X-1]{pottschmidt:98a}.

\section*{Variability of Spectral and Temporal Properties}

Fourier frequency dependent time lags of up to $\sim$0.1\,sec are known to
exist between different energy bands in \mbox{Cyg X-1}. While the lags
increase with energy, they cannot be explained by the diffusion time scale
of photons in a Compton corona alone \citep{miyamoto:89a,nowak:98a}.
Nevertheless, the characteristic time lag ``shelves'' allow to roughly
constrain coronal parameters \citep{nowak:98c}. We find that over the
course of weeks, the typical time lag in the hard state can vary by at
least a factor three (Fig.~\ref{fig:zrmcompare}, left panel). The first
three observations show a gradual decrease in the time lags, while the
fourth observation has intermediate values. This systematic development is
mirrored by the shot relaxation time scale $\tau$, which gets larger for
observations with smaller time lag (Fig.~\ref{fig:zrmcompare}, right
panel).

\begin{figure}
\centerline{\includegraphics[width=\figurewidth]{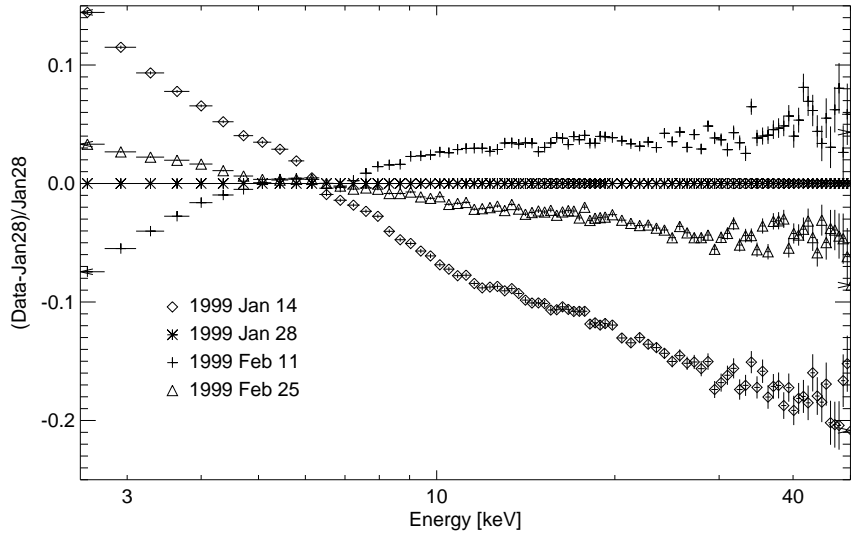}}
\caption{Relative deviation of the shape of the RXTE PCA count rate spectra
  from the observation of 1999 Jan 28.}
\label{fig:division}
\end{figure}

At the same time, the X-ray spectrum also changes systematically
(Fig.~\ref{fig:division}). Spectral fitting of black hole candidate spectra
with the PCA is severely affected by the uncertainty of the PCA response
matrix. Although the data exhibit a clear and varying hardening above
$\sim 10$\,keV, it is difficult to associate these changes with physically
interpretable spectral parameters. For example, both, a broken power-law
and a power law reflected from cold matter result in acceptable fits. This
behavior is similar to GX~339$-$4 \citep{wilms:98c}. In order to
characterize the spectral variability of \mbox{Cyg X-1} independently of
any spectral model, therefore, we directly compared the data in detector
space. Fig.~\ref{fig:division} displays the relative deviation of the four
observations with respect to the observation of 1999 Jan 28. \mbox{Cyg
  X-1} is clearly spectrally variable on a time basis of 14\,d (note that
part of the variation could be due to orbital modulation).

Comparison of Figs.~\ref{fig:zrmcompare} and~\ref{fig:division} shows that
a spectral hardening of the source correlates with a decrease of the time
lags and with an increase of the relaxation time $\tau$. Recently,
\citet{gilfanov:99a} also analyzed several of the public RXTE observations
of \mbox{Cyg X-1}. They found a variability of the spectral hardness of the
same order as presented here and an increase of the PSD break frequency
with the reflection fraction. They also confirmed for \mbox{Cyg X-1} a
correlation between the intrinsic spectral slope and the reflection
fraction \citep{zdziarski:99a}, as well as a relationship between two
temporal parameters, namely the PSD ``break frequency'' and the PSD ``hump
frequency'' \citep{wijnands:99a,psaltis:99a}.

Due to the long time basis of the available RXTE data it is also possible
to compare observations that are widely spaced in time, e.g., the 1999
monitoring observations with an observation made more than two years
earlier, in 1996 Oct 23. The latter has previously been published in a
series of papers \citep{dove:97c,nowak:98a,nowak:98c}. It was performed
shortly after the soft state of 1996, and we cautioned, therefore, that the
observation might still have been ``contaminated'' by soft state
peculiarities. But, the comparison with the observation of 1999 Feb 25
shows almost identical PSDs and time lags. So, we see that the source really
was in its hard state and that the hard state timing properties can be
reproduced with great accuracy on the time scale of years.

\section*{Discussion}

We have presented first results from our systematic analysis of RXTE data
of Cyg~X-1 in the hard state. Apparently, during the canonical hard state
this source can vary by up to a factor of $\sim 2$ in 2--50\,keV flux and by
up to a factor of three in the associated time lags within a few weeks. On
the other hand, we were also able to identify data with almost identical
spectral and temporal behavior spaced by more than two years.

As we noted in the previous section, there is possible evidence for a
correlation of the changes in the spectral and temporal behavior of the
source. Harder spectra appear to be associated with shorter time lags,
similar to the hard state of GX~339$-$4 \citep{nowak:99a}. A possible
interpretation would be that the accretion disk penetrates to smaller disk
radii at times of harder flux, thereby increasing the reflection fraction
of the Comptonized radiation \citep[see also][]{gilfanov:99a}, i.e.,
hardening the spectrum, and shortening the time-delay of the harder photons
(with the smaller system geometry corresponding to shorter lags).
Alternatively, the harder spectra might be due to changes in the coronal
parameters: our results might indicate that coronae with larger optical
depth and/or temperature are physically smaller. This is also consistent
with the development of the shot time scale in the sense that more
scattering events lead to longer relaxation times.

\subsection*{Acknowledgments}
We thank all participants in the 1999 broad band campaign for their
continued effort to obtain simultaneous radio through X-ray observations of
Cygnus~X-1. This work has been partially financed by DFG grant Sta~173/22
and a travel grant by the Deutsche Forschungsgemeinschaft to JW.

\bibliographystyle{jwapjbib}
\bibliography{mnemonic,jw_abbrv,apj_abbrv,/home/wilms/lib/bibtex/diplom,/home/wilms/lib/bibtex/agn,/home/wilms/lib/bibtex/ns,/home/wilms/lib/bibtex/accret,/home/wilms/lib/bibtex/bhc,/home/wilms/lib/bibtex/inst,/home/wilms/lib/bibtex/conferences}

\end{document}